\documentclass[prb,preprint]{revtex4-2} 

\usepackage{amsmath}  
\usepackage{amsfonts} 
\usepackage{graphicx} 

\newtheorem{Remark}{Remark}

\begin{document}


\title{Thermodynamic identities and thermodynamic consistency of Equations of State}

\author{Saad  Benjelloun}
\affiliation{MSDA, Mohammed VI Polytechnic University, Benguerir, Morocco}
\affiliation{CMLA, ENS-Paris Saclay, France}



\date{\today}

\begin{abstract}
We present a systematic approach to construct complete equations of state (EOSs), or to ensure thermodynamic consistency of complete and incomplete forms of EOSs using a minimal and sufficient set of relations. We apply the proposed approach to some common classical equations of state for pure materials. In fact, classical equations of state come generally in an incomplete form that hides essencial properties necessary for thermodynamic consistency. If not aware of such constraints one may generalize the EOS, or fit its thermodynamic parameters from emprirical data or from ab-initio models in an inconsistent way. 
\end{abstract}

\maketitle 

\section{Introduction}\label{S:1:intro}
When a  system is at thermodynamic equilibrium under a given set of conditions, it is said to be in a definite thermodynamic state. The state of the system can be described by a number of state quantities, called state functions. In the present paper, we consider closed systems with no change in composition, and in this case only two of these state quantities are independent. Hence, the state functions can be linked either by functional relations (e.g. an equations of state) which specifies a closed formula relationships between these quantities, or by thermodynamical identities linking the infinitesimal variations (or derivatives) for each triplet of these state quantities. If we limit ourselves to the $5$ fundamental state functions : $\rho =\frac{1}{v}$, $p$, $T$, $s$, and $e$, being respectively the specific density (the inverse of the specific volume), pressure, temperature, specific entropy and specific internal energy, then $C^3_5  =10$ thermodynamical identities can be expressed for a closed system. Other state functions such as specific enthalpy, specific free energy and specific Gibbs energy are obtained as simple functions of these fundamental quantities (e.g. $h = e+\frac{p}{\rho}$, $A  = e - Ts $, $g = e+ \frac{p}{\rho} -Ts $). Some authors may refer to $p$, $v$ and $T$, as state variables for being quantities directly accessible using measure instruments, here we will use the term state function or state variable interchangebly.

If we choose $\rho$ and $s$ as independant thermodynamic variables, the variations of the other three fundamental variables may be expressed throught the three identities : 

\begin{equation}
dp \, = \, c^2 \, d\rho+\rho k \,T \, ds \label{RSP}.
\end{equation}

\begin{equation}
de=Tds + \frac p{\rho^2}d\rho\label{RSE}.
\end{equation}

\begin{equation}
dT=\frac{T}{C_v}ds+\frac{kT}\rho d\rho \label{RST}.
\end{equation}

where we introduced the thermodynamic coefficients : $c^2$ the square of the speed of sound, $k$ the Gruneisen parameter and $C_v$ the thermal capacity at constant volume. 
\begin{Remark}
We assume here that the choice of $(\rho,s)$ as independant variables is possible. In general when adopting a given EOS for a material not all couple of variables are independants. For example, for perfect gazes, on can not choose $e$ and $T$ as independent variables. 
\end{Remark}

An equation of state (EOS) given in form $e= e(\rho,s)$ is said to be a complete EOS, because all the other state functions can be computed from it. $T(\rho,s) = \frac{\partial e}{\partial s}$,  $p(\rho,s) = \frac{\partial e}{\partial s}$. An EOS of type $p=p(\rho,T)$ is not complete. See remark \ref{remPVT}.

The quantities $c^2$, $k$ and $C_v$ are first order quantities in the sens that they are defined as derivatives (or coefficients) of the fundamental state functions (e.g. $c^2 = \frac{\partial p}{\partial \rho} )_s$), and are also functions of the state, so of the chosen independant variables $(\rho, s)$. Identities \eqref{RSP}-\eqref{RST} implies that Shwartz-type relations should be verified, so we have from \eqref{RSP} :
 
\begin{equation}
\left.\partial_s c^2  \right)_{\rho}= \left. \partial_{\rho} (\rho \, k \, T \right)_s  \label{RSconst1} 
\end{equation}

And from \eqref{RSE} we should have  $ \left.\partial_{\rho}T \right)_s = \left.\frac{1}{\rho^2}\partial_{s} p\right)_{\rho} = \frac{kT}{\rho}$ (1st Maxwell relation) and this is already satisfied through \eqref{RST} and \eqref{RSP}. Finally from \eqref{RST} we need to have :

\begin{equation}
\partial_{\rho} \left( \frac{T}{C_v} \right) _s  = \partial_{s} \left(\frac{kT}\rho\right)_{\rho} \label{RSconst2}
\end{equation}

Equations \eqref{RSconst1} and \eqref{RSconst2} can be simplified to the form : 

\begin{equation}
\partial_{s} \left( c^2 \right)_{\rho} = kT(k+1) + \rho T \partial_{\rho} \left( k \right)_{s}  \label{RSconst1s} 
\end{equation}

\begin{equation}
\partial_{\rho} \left( C_v \right)_{s} = -\frac{C_v {} ^2}{\rho} \partial_{s} \left( k \right)_{\rho}  \label{RSconst2s} 
\end{equation}


Hence, we see that in ($\rho$, $s$) variables and up to a reference state ($\rho_0$, $s_0$, $p_0$, $T_0$, $e_0$), a complete equation of state may be given by any three thermodynamic parameters functions $c^2 (\rho,s)$, $k(\rho,s)$ and $C_v(\rho,s)$ as long as they are verifiying the constraints \eqref{RSconst1s} and \eqref{RSconst2s}. We will refer to this choice as a $\{(\rho, s), c^2, C_v, k\}$ representation of the EOS.

\section{Thermodynamic identities and thermodynamic relations} \label{sec:Thermo_ID}

From \eqref{RSP}-\eqref{RST} one can derive the other 7 identities involving triplets of the 5 fundamental thermodynamic functions, to complete the set of the  $C^3_5 = 10$ possible identities :

\begin{equation}de=\frac{1}{\rho k}dp+\frac{1}{\rho^2} \left(p - \frac{\rho c^2}{k} \right)d\rho\label{RPE}.\end{equation}
\begin{equation}dT=\frac{1}{\rho k C_v}dp+\frac{1}{\rho} \left(  kT - \frac{c^2}{k C_v} \right) d\rho \label{RPT}.\end{equation}
\begin{equation}dT=\frac1{C_v}de+\left(kT-\frac p{\rho C_v}\right)\frac{d\rho}{\rho}\label{RET},\end{equation}
\begin{equation}de=\frac{p}{\rho^2 c^2} dp+\left(T - \frac{pkT}{\rho c^2}\right) ds  \label{SPE},\end{equation}
\begin{equation}dT=\frac{kT}{\rho c^2} dp+ \left( \frac{c^2 - k^2 T C_v }{C_v c^2}\right) T ds  \label{SPT},\end{equation}
\begin{equation}dT=\frac{\rho kT}{ p} de+ \left( \frac{p -\rho k  C_v T }{pC_v }\right) T ds  \label{SET},\end{equation}
\begin{equation}dT=\frac{\rho c^2  - \rho k^2 T C_v}{ C_v (\rho c^2 - pk) } de+ \frac{\rho k  C_v T - p }{\rho C_v (\rho c^2 - pk ) } dp  \label{PET },\end{equation}

Thermodynamic identities relating `secondary' state functions $h$, $A$ and $g$ to any two primary variables among $(\rho, p, e, s, T)$ can also be derived. For instance for $h(p,\rho)$  :
$$
dh = d (e+\frac{p}{\rho}) = \frac{1}{\rho k}dp+\frac{1}{\rho^2} \left(p - \frac{\rho c^2}{k} \right)d\rho + \frac{1}{\rho}dp - \frac{p}{\rho^2} d\rho
$$ 
$$
= \frac{1+k}{\rho k} dp -\frac{\rho \, c^2}{k} d\rho
$$

We may retrieve from \eqref{RPE} the usual definition of $k$ as Gruneisen parameter $ k = \left. \frac{1}{\rho} \frac{\partial p} {\partial e}\right)_{\rho}$. Moreover, expressions for all the other thermodynamic coefficients may be derived from the identities above as functions of $c^2$, $C_v$ and $k$ (we recall that these three coefficients are linked by \eqref{RSconst1s} and \eqref{RSconst2s} ):

\begin{itemize}
\item Isothermal bulk modulus $K_T$, Isothermal compressibility $\beta_T = \frac{1}{K_T}$, and isothermal speed of sound (for example from \eqref{RPT}):
$$K_T = \left. \rho \frac{\partial p}{\partial \rho}\right)_T = \rho \left( c^2 - k^2 C_v T \right)$$ 
$$\beta_T = \frac{1}{\rho c^2 - \rho k^2 C_v T},  \, \,   c_T^2 =c^2 -  k^2 C_v T. $$
Isentropic bulk and compressibility coefficients are given by $K_s = \rho c^2$ and $\beta_s = \frac{1}{K_s}$.
\item Isobaric thermal expansion coefficient $\alpha_p$ and isochoric thermal expansion coefficient $\alpha_v$ :
$$\alpha_p = - \frac{1}{\rho} \left( \frac{\partial \rho}{\partial T} \right)_p = \frac{k C_v}{c^2 - k^2 T C_v}$$.
$$\alpha_v =\left. \frac{\partial P}{\partial  T} \right)_{\rho} = \rho k C_v $$ 
One can also define an isentropic thermal expansion coefficient : $$ \alpha_s = - \frac{1}{kT} = - \left. \frac{1}{\rho} \frac{\partial \rho}{\partial T}\right)_s$$

\item Heat capacity coefficients : Isobaric heat capacity $C_p$, and heat ratio $\gamma$.
 $$C_p = \left. T \frac{\partial s}{\partial T} \right)_p = \frac{C_v c^2}{c^2 - k^2 T C_v } \, ,\,\, \gamma = \frac{c^2}{c^2 - k^2 T C_v }.$$

\end{itemize}

Also we note that, so far, the second principle of thermodynamics is not taken into account. We will briefly address this in section \ref{2dPrinciple}.
 

As noted an EOS of the form $e = e(\rho,s)$ (or $s=s(\rho,e)$, or $\rho=\rho(e,s)$ ) is complete as $p(\rho,s)$ and $T(\rho,s)$ can be computed from it.
Alternatively the set of equations \eqref{RSconst1s}-\eqref{RSconst2s} provide a systematic way to check the thermodynamic consistency (in regard to definitions and the first principle) of an EOS in $\rho$, $s$ variables. It provides also a systematic way to construct a complete EOS either from empirical relations or ab-initio considerations. The choice of the representation $\{ (\rho , s), c^2, C_v, k\} $ is arbitrary and depending on the context, one can use other representations. We will present here few other choices that will be applied to several examples of equations of state.

 \section{Different presentations for thermodynamic states description}
\subsection{Representation :  $\{( \rho , T) , {\alpha_v, c_T^2,  C_v}\} $ }\label{subsectionRT}


In this presentation we suppose  $(\rho, T)$  is a valid couple of independant variables for the thermodynamic evolution of the material. We express the different coeffcients as functions of  $\{\alpha_v, c_T^2,  C_v\}$ :

\begin{equation} dp  = c_T^2 d\rho + \alpha_v dT \, , \end{equation}
\begin{equation} de =\frac {p - \alpha_v T }{\rho} \frac{d\rho}{\rho} +  C_v  dT \, ,\end{equation}
\begin{equation} ds = \frac{C_v}{T} dT - \frac{\alpha_v}{\rho^2} d\rho \end{equation}
and the consistency relations are given by :  
\begin{equation} \label{beta}
\partial_T(c_T^2 )_{\rho} =  \partial_{\rho}(\alpha_v )_{T}
 \end{equation}
\begin{equation} \label{Cv}
\frac{1}{T} \partial_{\rho}(C_v )_T =  - \frac{1}{\rho^2}\partial_{T}(\alpha_v)_{\rho}
 \end{equation}
 
 The other thermodynamic parameters can be given in function of $\alpha_v$, $c_T^2$ and $C_v$ by :
 \begin{equation}
 \alpha_p = \frac{\alpha_v}{\rho c_T^2}  \, , \, \,  k=\frac{\alpha_v}{\rho C_v}   \label{paramRT1}
 \end{equation}
 
 \begin{equation}
 c^2 = c_T^2 + \frac{\alpha_v^2 T}{\rho C_v} \, , \, \, \gamma = 1+ \frac{\alpha_v^2 T}{\rho C_v c_T^2}  
 \end{equation}
 \begin{equation}
 C_p = C_v+ \frac{\alpha_v^2 T}{\rho c_T^2} \label{paramRT5}
\end{equation}
 
\begin{Remark}\label{remPVT}
A PVT relation of the form $p=f(\rho,T)$ sets $\alpha_v$ and $c_T^2$ (with relation \eqref{beta} satisfied). The $C_v(\rho,T) $ parameter is then to be defined to form a complete EOS, with the constraint \eqref{Cv}; Hence, $C_v(\rho, T)$ should be defined only along an isochore :  $T \rightarrow C_v(\rho_0, T)$. 
\end{Remark}
 
 \begin{Remark}\label{potentiel}
 The Helmotz free energy is defined by :
 $$
 A(\rho,T) = e(\rho,T) - T s(\rho,T)
 $$
and if the form $A(\rho,T)$ is given all the thermodynamic quantities can be found from $A$ by the following relations (in this order) :
 $s = - \left. \frac{\partial A}{\partial T} \right)_\rho$, $ e = A +Ts $, $ p = \rho^2 \left. \frac{\partial A}{\partial \rho} \right)_T $ , as we have $dA = \frac{p}{\rho^2} d\rho - s dT$. 
 
From $s(\rho,T)$ we can compute $C_v$ and $\alpha_v$ and from $p(\rho,T)$ we can compute $c_T^2$. Of course, this approach will automatically satisfy the relations of type \eqref{beta} and \eqref{Cv}.
 
On other hand giving a PVT relation $p = p(\rho,T)$, one sets only $\left. \frac{\partial A}{\partial \rho} \right)_T$, so $A$ is set up to a function $A_0 = A(\rho_0, T) = \int_{T_0}^{T} s(\rho_0, T) dT + e_0 - T_0 s_0 $ of $T$ :
$$A(\rho,T) =  \int_{\rho_0}^{\rho} \frac{p(r, T)}{r^2} dr + A_0(T)$$ 
\end{Remark}

 
\subsection{Representation :  $\{( \rho , p) , {c^2, \alpha_v = \rho k C_v, k}\} $ }

We have :

\begin{equation}dT=\frac{1}{\alpha_v}dp -  \left( \frac{c^2 }{\alpha_v}  - \frac{ k T}{\rho}  \right)  d\rho\end{equation}
\begin{equation}de=\frac{1}{\rho k}dp+\frac{1}{\rho} (\frac{p}{\rho} - \frac{c^2}{k} )d\rho \end{equation}
\begin{equation}ds = \frac{1}{\rho kT} dp - \frac{c^2}{\rho kT} d\rho \end{equation}

With the Shwartz consistency relations :
\begin{equation}\label{PRconst1}
\rho c^2  \frac{\partial k}{\partial p}  - \rho k \frac{\partial c^2}{\partial p} + \rho \frac{\partial k}{\partial \rho}  + k^2 + k  = 0 
\end{equation}
\begin{equation}\label{PRconst2}
\frac{C_{v}^{2}}{\rho}\frac{\partial k}{\partial p} +\frac{c^2}{ \rho k T } \frac{\partial  C_{v}}{\partial p} + \frac{1}{ \rho k T }\frac{\partial  C_{v}}{\partial \rho}=0
\end{equation}

We note that the last identity is the same as \eqref{RSconst2s}.
The other thermodynamic coefficients can be expressed with relations presented previously in the $(\rho,s, c^2 ,C_v, k)$ representation.

\subsection{Representation :  $\{(p, T) , { C_p, \kappa_T = \frac{1}{\rho^2 c_T^2}, \eta =\frac{\alpha_p}{\rho} \}} $ }\label{subsectionpT}

We can write :
\begin{equation} dv = d (1/\rho)  = - \kappa_T dp  + \eta dT \label{RPT_tiers}.\end{equation}
\begin{equation} de =  \left( C_p  - p \eta \right) dT - (\eta T - p \kappa_T) dp  \label{PET_tiers},\end{equation}
\begin{equation}ds = \frac{C_p}{T} dT - \eta dp    \label{SPT_tiers},\end{equation}

with the Shwartz relations :
$$
\frac{\partial \kappa_T}{\partial T} + \frac{\partial \eta}{\partial p} = 0
$$
$$
\frac{\partial \eta}{\partial T} + \frac{1}{T}\frac{\partial C_p}{\partial p} = 0
$$

The other thermodynamic coefficients can be expressed as :
$$
k= \frac{\alpha_p c_T^2}{C_p - T \alpha_p^2 c_T^2}
$$
$$
\alpha_v = \rho c_T^2 \alpha_p 
$$
$$
C_v =  C_p -  T \alpha_p^2 c_T^2 \, , \, \, \gamma = \frac{C_p}{ C_p - T \alpha_p^2 c_T^2}
$$
$$
c^2 = \frac{C_p c_T^2}{ C_p - T \alpha_p^2 c_T^2}
$$

\section{Link with classical thermodynamic relations}
Obviously, whatever is the chosen representation, the expressions for thermodynamic parameters verify all known thermodynamic relations:
$$ \beta_T = \frac{1}{\rho c_T^2} = \frac{\gamma}{\rho c^2} = \gamma \beta_s\,, \, \, K_T= \rho c_T^2 = \frac{\rho c^2}{\gamma}  =\frac{K_s}{\gamma}  ~~ \text{(Reech relation)} $$
$$\text{with} ~\gamma = \frac{C_p}{C_v} = \frac{K_s}{K_T}= \frac{c^2}{c_T^2} = \frac{c^2}{c^2 - k^2 T C_v },$$
$$c = \sqrt{\frac{K_s}{\rho}} = \sqrt{\frac{\gamma K_T}{ \rho}}      ~~ \text{(Newton-Laplace)} $$
$$\alpha_p = \rho k C_v \beta_T =  \rho k C_p \beta_s = \frac{k C_p}{ c^2} ~~ \text{(Gruneissen relation)} $$
$$C_p - C_v =  T\frac{\alpha^2}{\rho \beta_T}  ~~ \text{(Mayer relation)},$$
and 
$$\beta_T - \beta_s =  \frac{\alpha^2 T}{\rho C_p}  ~~ \text{(Mayer relation)},$$

The advantage of the presented approach, where a representation is chosen to work with, is being systematic and avoiding wrangling with the above famous relations to express one coefficient as function of (two/three) others. It also does not ommit the consistency relations between these coeffiscents derivatives when they are defined by empiric relations, fitting experimental data or from ab-initio considerations. 

We show in the next section how this systematic approach can be applied to introduce well know Equations of State along with their consistency conditions that are usually omitted in textbooks or presented as a complex results of thermodynamic relation wrangling.

\section{A revisit of some classical Equations Of State}\label{S:1:EOS1phase}

Equations of state are useful in describing the properties of fluids, mixtures of fluids, solids, and other forms of matter. Many examples of EOS are adopted  to relate the different state functions of a given material, in diferent areas of physical sicences (fluids, mechanics, geology, astrophysics...etc).

\subsection{Ideal and Perfect gaz EOS}
The Ideal gas EOS is given by  
\begin{equation}\label{pvt_ideal_gaz}
p = \rho r T.
\end{equation}

Hence and if we use the $\{ (\rho,T), \alpha_v, c_T^2, C_v\}$ representation presented in section \ref{subsectionRT} we get :
\begin{equation}
\alpha_v = r \, \rho
\end{equation}

\begin{equation}
c_T^2 = r T
\end{equation}

with relation \eqref{beta} satisfied. Equation \eqref{Cv} implied then that $C_v$ is function of $T$ only. So, we conclude that a complete Ideal Gaz EOS is completely defined by a function $T \rightarrow C_v(T)$ and the PVT relation \eqref{pvt_ideal_gaz}. All the thermodynamic coefficients can be calculated throught equations \eqref{paramRT1}-\eqref{paramRT5}. 

$$
k(\rho, T) = \frac{r}{C_v(T)}\, ,\,\,  \alpha_p = \frac{1}{ T },
$$ 
$$
c^2 = rT  \left(1+\frac{r}{C_v(T)}\right),
$$
$$
\gamma = \frac{c^2}{c_T^2} = 1 + \frac{r}{C_v(T)}, 
$$
$$
C_p = C_v(T) +  r.
$$

Usually the $C_v$ dependence on $T$  only is given in textbooks through proving the dependences of $e$ or $h$ on $T$ only and this is obtained by wrangling the thermodynamic relations (Maxwell relations) and identities.

Coming back to the EOS, and knowing a refrence state $(\rho_0, T_0, p_0, e_0, s_0)$ the expression of other state functions can be derived by integrating the thermodynamic identities \ref{sec:Thermo_ID}:
$$
p(\rho, T) = r\rho T
$$
$$
e(T) = \int_{T_0}^T C_v (t) dt + e_0
$$
 
$$
s (\rho, T) = - r  \, ln(\frac{\rho}{\rho_0}) +  \int_{T_0}^T \frac{C_v (t)}{t} dt + s_0
$$
The Helmotz free energy is given by :
$$ A(\rho,T) = e -Ts =  \int_{T_0}^T C_v (t) dt - T \int_{T_0}^T \frac{C_v (t)}{t} dt  +  r. T . ln(\frac{\rho}{\rho_0}) + e_0 - T s_0   $$

These formulae simplifies further if $C_v$ is supposed constant, which is called by some authors the Perfect Gaz EOS.
$$p=r\rho T ~~ , ~~~~e = C_v(T-T_0) + e_0 ~~,~~~~ s =-r\, ln(\frac{\rho}{\rho_0}) +  C_vln(\frac{T}{T_0}) + s_0
 $$, 

We note that in this case, one can not chose $T_0 =0$ or $\rho_0 =0$ as $\frac{1}{T}$ and $\frac{1}{\rho}$  needs to be integrable at $T_0$ and $\rho_0$. Hence the Perfect Gaz model cannot be valid around these conditions (absolute zero and infinite volume). Other ab-initio models for $C_v$ with   $C_v(T) \approx T^3$ (e.g. Debye model ) or $C_v(T) \approx T$ near $T=0$, are consistent with the Ideal Gaz EOS.


\subsection{Other form for Perfect gaz EOS}

Some authors introduce the perfect gaz EOS with the relation 
\begin{equation}\label{prp_perfect_gaz}
p = \rho (\Gamma - 1) e
\end{equation}

with a constant parameter $\Gamma$.

It is important to note that this form is indeed consistent with \eqref{pvt_ideal_gaz} only and only if the two conditions hold 
$$
\begin{cases}
C_v \text{ is constant} \\
\text{There is a reference state such as } e_0= C_v \,T_0
\end{cases}
$$

If any one of the conditions does not hold then the form \eqref{prp_perfect_gaz} is rather a special form of Mie-Gruneissen EOS, for witch the Gruneissen parameter is given by $k = \Gamma- 1$ and for which a varying $C_v$ function is adminissible only if it depends on $s$ only, as we will see in section \ref{MG}. To stress it more one cannot adopt \ref{prp_perfect_gaz} along with a correlation  $C_v = C_v(T)$.

\subsection{Van de Waals EOS}

The Van der Waals EOS is given by a pvt relation :
$$
p = \frac{\rho rT}{1 - b\rho} - a\rho^2
$$
 We have from the PVT relation 


 $$
 c_T^2  = \frac{rT}{(1-b\rho)^2}  - 2 a \rho \, ,\,\, \alpha_v = \frac{r \rho}{1-b\rho}
 $$
 and the relation \eqref{Cv} implies :
 $$
 \partial_{\rho} C_v )_T = 0
 $$
 
As for the perfect gaz, the VDW EOS is completely defined by a  $T \rightarrow C_v(T)$ function. Having a model $C_v(T)$ we can compute the other thermodynamic properties as functions of $\rho$ and $T$.
$$
k(\rho, T) = \frac{r}{1-b\rho}.\frac{1}{C_v(T)}\, ,\,\,  \alpha_p = \frac{r(1-b\rho)}{ r T - 2 a \rho (1-b\rho)^2},
$$ 
$$
c^2 = \frac{rT}{(1-b\rho)^2}  \left(1+\frac{r}{C_v(T)}\right) - 2 a \rho,
$$
$$
\gamma = \frac{c^2}{c_T^2} = 1 + \frac{ r^2T}{ rT  - 2 a \rho (1-b\rho)^2  }.\frac{1}{C_v(T)}, 
$$
$$
C_p = C_v(T) + \frac{ r ^2T  }{rT  - 2 a \rho (1-b\rho)^2 }.
$$

Also, given a reference state, we have the other state functions given by :

 $$e = e_0 - a(\rho - \rho_0) + \int_{T_0}^T C_v(t) dt$$

$$s = s_0 -\int_{\rho_0}^{\rho} \frac{r}{x(1-bx)} dx  + \int_{T_0}^T \frac{C_v(t)}{t} dt$$
$$
=  s_0 + r . ln( \frac{1/\rho-b}{1/\rho_0 -b}) + \int_{T_0}^T \frac{C_v(t)}{t} dt
$$
and for the free energy :
$$
A(\rho,T) = A_0 - a(\rho - \rho_0) - r T ln(\frac{1/\rho-b}{1/\rho_0 -b})  + \int_{T_0}^T C_v(t) dt   - T \int_{T_0}^T \frac{C_v(t)}{t} dt
$$ 

We note that here that in contrast of the ideal gaz, a constant $C_v$ model (which is compatible with VDW EOS) does not imply necessarly a contant $\gamma$ or a constant $C_p$. A constant $C_v$ with VDW EOS implies in fact a special case of  Mie-Gruneisen model $k=k(\rho)$ presented in the section \ref{MG}.

\subsection{Redlich-Kwong EOS}\label{R-K}

The RK EOS is given by :
$$
p = \frac{\rho rT}{1 - b\rho} - \frac{a\rho^2}{T^{\sigma}(1+b\rho)}
$$

where $a$, $b$, $r$ and ${\sigma}$ are constant parameters. From the PVT relation :
we get 
 $$
 c_T^2  = \frac{rT}{(1-b\rho)^2}  -  a \rho \frac{2+\rho b}{(1+\rho b)^2} \, ,\,\, \alpha_v = \frac{r \rho}{1-b\rho} + \frac{a \rho^2 {\sigma} }{T^{{\sigma}+1} (1+b\rho)}
 $$
 and the relation \eqref{Cv} implies :
 $$
 \partial_{\rho} C_v )_T =  \frac{a {\sigma} ({\sigma}+1)}{(1+b\rho)} \frac1{T^{{\sigma}+1}} 
 $$
 Hence we have a special dependence of $C_v$ on $\rho$ :
 $$
 C_v(\rho,T) =  \frac{a {\sigma} ({\sigma}+1) ln(1+b\rho)}{b T^{{\sigma}+1}}  + \phi(T)
 $$
 
The R-K EOS is completely defined by the function  $\phi(T)$ function. A $T$ only dependent $C_v$ is NOT compatible with the K-S equation state unless $a=0$, $\sigma=0$ or  $\sigma=-1$. Having a model $C_v(\rho,T)$ we can compute the other thermodynamic properties as functions of $\rho$ and $T$.

$$k(\rho, T) = \frac{r}{(1-b\rho) C_v} + \frac{a \rho {\sigma}}{T^{{\sigma}+1} (1+b\rho) C_v},$$
$$  \alpha_p =\frac{ \frac{r}{1-b\rho} + \frac{a \rho {\sigma}}{T^{{\sigma}+1} (1+b\rho)} }{\frac{rT}{(1-b\rho)^2}  -  a \rho \frac{2+\rho b}{(1+\rho b)^2}}$$ 
$$c^2 = c_T^2 + \frac{\alpha_v^2 T}{\rho C_v} \, , \,\, \gamma = \frac{c^2}{c_T^2}\, ,\,\, C_p = C_v + \frac{\alpha_v^2 T}{\rho c_T^2}.$$

Also, given a reference state, and from
$$
de = -\frac{a({\sigma}+1)}{T^{\sigma}(1+\rho b)} d\rho +  \left[ \frac{a {\sigma} ({\sigma}+1) ln(1+b\rho)}{b T^{{\sigma}+1}}  + \phi(T) \right] dT
$$
we have :
$$e = e_0 + \frac{a({\sigma}+1)}{b T^{\sigma}} ln( \frac{b\rho_0 + 1}{b\rho + 1} ) - \frac{a ({\sigma}+1) ln(1+b\rho_0)}{b} \left[\frac{1}{T^{\sigma}} - \frac{1}{T_0^{\sigma}} \right] + \int_{T_0}^T \phi(t) dt$$
$$ = e_0  -\frac{a ({\sigma}+1) }{b}  \left[\frac{ln(1+b\rho)}{T^{\sigma}} - \frac{ln(1+b\rho_0)}{T_0^{\sigma}} \right]+ \int_{T_0}^T \phi(t) dt$$
And for the entropy identity :
$$ds =  \left[ \frac{a {\sigma} ({\sigma}+1) ln(1+b\rho)}{b T^{{\sigma}+2}}  + \frac{\phi(T)}{T} \right] dT -  \left[ \frac{r }{\rho(1-b\rho)} + \frac{a {\sigma}}{T^{{\sigma}+1} (1+b\rho)} \right] d\rho$$
$$s = s_0 -\int_{\rho_0}^{\rho} \frac{r}{x(1-bx)} dx  + \int_{T_0}^T \frac{\phi(t)}{t} dt + \frac{a {\sigma}}{b T^{{\sigma}+1}}ln(\frac{ 1+b\rho_0}{ 1+b\rho}) - \frac{a {\sigma} ln(1+b\rho_0)}{b} \left[ \frac{1}{T^{{\sigma}+1}} - \frac{1}{T_0^{{\sigma}+1}} \right] $$
$$
= s_0 +  r . ln( \frac{1/\rho-b}{1/\rho_0 -b})  + \int_{T_0}^T \frac{\phi(t)}{t} dt - \frac{a {\sigma}}{b} \left[ \frac{ ln(1+b\rho)}{T^{{\sigma}+1}} - \frac{ln(1+b\rho_0)}{T_0^{{\sigma}+1}}\right]
$$
So for the free Helmotz energy :
\begin{multline}
A(\rho,T) = A_0 - r T ln(\frac{1/\rho-b}{1/\rho_0 -b})   -\frac{a ({\sigma}+1) }{b}  \left[\frac{ln(1+b\rho)}{T^{\sigma}} - \frac{ln(1+b\rho_0)}{T_0^{\sigma}} \right]
\\
+ T \frac{a {\sigma}}{b} \left[ \frac{ ln(1+b\rho)}{T^{{\sigma}+1}} - \frac{ln(1+b\rho_0)}{T_0^{{\sigma}+1}}\right]
\\
+\int_{T_0}^T\phi(t) dt   - T \int_{T_0}^T \frac{\phi(t)}{t} dt
\end{multline} 
\begin{multline}
A(\rho,T) = A_0 - r T ln(\frac{1/\rho-b}{1/\rho_0 -b})   - \frac{a }{b}  \frac{ln(1+b\rho)}{T^{\sigma}}  
\\
 +\frac{a ({\sigma}+1) }{b}  \frac{ln(1+b\rho_0)}{T_0^{\sigma}}  - T \frac{a {\sigma}}{b} \frac{ln(1+b\rho_0)}{T_0^{{\sigma}+1}}
\\
+\int_{T_0}^T\phi(t) dt   - T \int_{T_0}^T \frac{\phi(t)}{t} dt
\end{multline} 

\begin{multline}
A(\rho,T) = A_0 - r T ln(\frac{1/\rho-b}{1/\rho_0 -b})   - \frac{a }{b}  \frac{ln(1+b\rho)}{T^{\sigma}} 
\\
+ \frac{a }{b}  \left( {\sigma}+1 -  \frac{T}{T_0} {\sigma} \right) \frac{ln(1+b\rho_0)}{T_0^{{\sigma}}}
\\
+\int_{T_0}^T\phi(t) dt   - T \int_{T_0}^T \frac{\phi(t)}{t} dt
\end{multline}

The formulae we give here for $A(\rho,T)$, and for $C_p$ for example, are more explicit than the one given in \cite{Matsumoto}, with a direct link with the $C_v(\rho, T)$ correlation (that can be fitted experimentally). 

\subsection{Peng Robinson}

\begin{align}
       p &= \frac{R\,T}{V_m - b} - \frac{a\,\alpha(T)}{V_m^2 + 2bV_m - b^2} \\[3pt]
       a &\approx 0.45724 \frac{R^2\,T_c^2}{p_c} \\[3pt]
       b &\approx 0.07780 \frac{R\,T_c}{p_c} \\[3pt]
  \alpha &= \left(1 + \kappa \left(1 - T_r^\frac{1}{2}\right)\right)^2 \\[3pt]
  \kappa &\approx 0.37464 + 1.54226\,\omega - 0.26992\,\omega^2 \\[3pt]
     T_r &= \frac{T}{T_c}
\end{align}
We rewrite the P.R. EOS with the specific quantities as :
$$
p = \frac{r\,T \rho}{1 - b\rho} - \frac{a(T) \rho^2}{1+ 2b \rho - b^2 \rho^2} 
$$
We can compute directly ($v=\frac{1}{\rho}$):
$$
c_T^2 = -\frac{rT}{(v - b)^2} + 2a(T) \frac{(v + b)}{[v^2 + 2b v - b^2]^2}\,, \,\,
\alpha_v = \frac{r}{v - b} - \frac{a'(T)}{v^2 + 2 bv - b^2}
$$
and this EOS imposes that : 
$$
\partial_{\rho} C_v = - T/\rho^2 \partial_T  \left[ \frac{r}{v - b} - \frac{a'(T)}{v^2 + 2 bv - b^2} \right]
$$
$$
\partial_{\rho} C_v =  \frac{Ta''(T)}{ 1 + 2 b\rho - b^2\rho^2}
$$
Hence
$$
C_v(\rho,T) =   \frac{Ta''(T)}{2 \sqrt{2} b} ln\left(\frac{\sqrt{2} -1 + b\rho}{\sqrt{2} + 1 - b\rho} \right)+ \phi(T)
$$

And for the other thermodynamic coefficient we have :
$$
k =\frac{\alpha_v}{\rho C_v} =  \frac{r v}{(v - b)C_v} - \frac{a'(T) v}{(v^2 + 2 bv - b^2)C_v}
$$
$$
c^2 =  c_T^2 + \frac{T}{C^v}  \left( \frac{r v}{(v - b)} - \frac{a'(T) v}{(v^2 + 2 bv - b^2)} \right)^2     
$$
$$
\gamma = 1 + \frac{T}{C^v}  \frac{\left( \frac{r v}{(v - b)} - \frac{a'(T) v}{(v^2 + 2 bv - b^2)} \right)^2}{-\frac{rT}{(v - b)^2} + 2a(T) \frac{(v + b)}{[v^2 + 2b v - b^2]^2}}    
$$
$$
C_p = C_v +  T \frac{\left( \frac{r v}{(v - b)} - \frac{a'(T) v}{(v^2 + 2 bv - b^2)} \right)^2}{-\frac{rT}{(v - b)^2} + 2a(T) \frac{(v + b)}{[v^2 + 2b v - b^2]^2}}   . 
$$
$$
C_p = \frac{Ta''(T)}{2 \sqrt{2} b} ln\left(\frac{\sqrt{2} -1 + b\rho}{\sqrt{2} + 1 - b\rho} \right)+ \phi(T) +  T \frac{\left( \frac{r v}{(v - b)} - \frac{a'(T) v}{(v^2 + 2 bv - b^2)} \right)^2}{-\frac{rT}{(v - b)^2} + 2a(T) \frac{(v + b)}{[v^2 + 2b v - b^2]^2}}.
$$

Hence for $e$ as function of $T$ and $\rho$ :
\begin{equation} de =\frac {p - \alpha_v T }{\rho} \frac{d\rho}{\rho} +  C_v  dT \, ,\end{equation}
\begin{equation} ds = \frac{C_v}{T} dT - \frac{\alpha_v}{\rho^2} d\rho \end{equation}

$$
de =  \frac{T a'(T) - a(T) }{1 + 2 b \rho - b^2 \rho^2}   d\rho +  \left[   \frac{Ta''(T)}{2 \sqrt{2} b} ln\left(\frac{\sqrt{2} -1 + b\rho}{\sqrt{2} + 1 - b\rho} \right)+ \phi(T) \right] dT
$$

$$ e(\rho,T) = e_0 + \frac{T a'(T) - a(T) }{2 \sqrt{2} b} ln\left(\frac{\sqrt{2} -1 + b\rho}{\sqrt{2} + 1 - b\rho} \right) + \int_{T_0}^T \phi(t) dt$$
And for the entropy :
$$ds =  \left[   \frac{a''(T)}{2 \sqrt{2} b} ln\left(\frac{\sqrt{2} -1 + b\rho}{\sqrt{2} + 1 - b\rho} \right) + \frac{\phi(T)}{T} \right] dT -  \left[  \frac{r}{\rho(1 - b\rho)} - \frac{a'(T)}{1 + 2 b\rho - b^2\rho^2} \right] d\rho$$
$$s = s_0 +  r . ln( \frac{1/\rho-b}{1/\rho_0 -b})  + \int_{T_0}^T \frac{\phi(t)}{t} dt - \frac1{2 \sqrt{2} b}\left[ a'(T) ln\left(\frac{\sqrt{2} -1 + b\rho}{\sqrt{2} + 1 - b\rho} \right) -   a'(T_0) ln\left(\frac{\sqrt{2} -1 + b\rho_0}{\sqrt{2} + 1 - b\rho_0} \right) \right]
$$
So for the free Helmotz energy :
\begin{multline}
A(\rho,T) = A_0 + \frac{T a'(T) - a(T) }{2 \sqrt{2} b} ln\left(\frac{\sqrt{2} -1 + b\rho}{\sqrt{2} + 1 - b\rho} \right) + \int_{T_0}^T \phi(t) dt   - T\int_{T_0}^T \frac{\phi(t)}{t} dt - \\  
r T . ln( \frac{1/\rho-b}{1/\rho_0 -b})  + \frac{T}{2 \sqrt{2} b}\left[ a'(T) ln\left(\frac{\sqrt{2} -1 + b\rho}{\sqrt{2} + 1 - b\rho} \right) -   a'(T_0) ln\left(\frac{\sqrt{2} -1 + b\rho_0}{\sqrt{2} + 1 - b\rho_0} \right) \right]
\end{multline} 


\subsection{Mie-Gruneissen}\label{MG}

All the EOS's presented before were in the form $P = f(\rho,T)$, hence we used a $(\rho, T)$ representation. The general Mie-Gruneissen EOS presented in this section (and the Stiffened gaz EOS of the next section) are given in the form $e(p,\rho)$ or $p(\rho,e)$.

The Mie-Gruneissen EOS is given by 
$$
p(\rho,e) = p_K(\rho) + \rho k(\rho) ( e - e_K(\rho)) 
$$
or in $(\rho, p)$ formulation
$$
e(p,\rho)  =   e_K(\rho) +  \frac{1}{\rho k(\rho)}  (p -  p_K(\rho))
$$

with two functions $e_K(\rho)$ and $p_K(\rho)$, satisfying the relation \cite{Houze}
$$\frac{d e_K(\rho)}{d \rho} = \frac{p_K(\rho)}{\rho^2}$$ so that $e_K$ and $p_k$ are defined along an isentrope $s=s_0$. 

We suppose  also that we know a reference state $(\rho_0, e_0, p_0, s_0, T_0)$, so that $p_K(\rho_0) = p_0 = p(e_0, \rho_0)$ and $e_K(\rho_0) = e_0 = e(\rho_0, p_0)$.

From the EOS and the identity $de = \frac{1}{\rho k} dp +  \frac{1}{\rho^2} (p - \frac{\rho c^2}{k}) d\rho$ we have then $k = k(\rho)$, and also $$c^2(p,\rho) = \left( \frac{k'}{k} + \frac{k+1}{\rho} \right) (p-p_K(\rho)) + p'_K(\rho)$$

A first consistency relation is automatically satisfied given the relation $e(p,\rho)$. 
Using \eqref{PRconst2} or \eqref{RSconst2}  we have :
$$
\partial_{\rho}(C^v)_{s} = 0
$$

Hence $C^v$ depends only on $s$ and then we have for the temperature 
$$
\frac{dT}{T} = \frac{1}{C^v(s)} ds + \frac{k(\rho)}{\rho} d\rho
$$

Hence to completely define the EOS in a $(\rho,s)$ presentation we are free to choose the two functions $k(\rho)$ et $C^v(s)$. We can then compute all the quantities in the representation $\rho,s$:  
$$
ln(T) = \int_{s_0}^s \frac{dS}{C^v(S)} + \int_{\rho_0}^{\rho} \frac{k(r)dr}{r} + ln(T_0)
$$

$$
T(\rho,s) = T_{0} e^{\int_{s_0}^s \frac{dS}{C^v(S)}} . e^{\int_{\rho_0}^{\rho} \frac{k(r)dr}{r}} = f(s) .\Theta(\rho)
$$

with $\Theta(\rho) = T_0 e^{\int_{\rho_0}^{\rho} \frac{k(r)dr}{r}} $ is the Debye temperature defined by $\frac{d ln(\Theta)}{d\rho} = \frac{k}{\rho}$, 
and $f(s) = e^{\int_{s_0}^s \frac{dS}{C^v(S)}}$.

Hence for the speed of sound we have :
$$
\partial_s(c^2)_{\rho} = \left[k(\rho) (k(\rho) +1) + \rho  k'(\rho)\right].\Theta(\rho).f(s)
$$
so
$$
c^2(\rho,s) = \left[k(\rho) (k(\rho) +1) + \rho  k'(\rho)\right].\Theta(\rho) \int_{s_0}^{s} f(\sigma) d\sigma + c_0^2(\rho)
$$
$$
= \left[k(\rho) (k(\rho) +1) + \rho  k'(\rho)\right]  \Theta(\rho)  F_{s_0}(s)+ c_0^2(\rho)
$$
$$
= \left[k(\rho) (k(\rho) +1) + \rho  k'(\rho)\right]  \Theta(\rho)  F_{s_0}(s)+ p_K'(\rho)
$$

For the pressure and from $dp = c^2 d\rho + \rho k T ds = c^2 d\rho + \rho k(\rho) f(s)\Theta(\rho) ds $ we get :
$$
p(\rho,s) = \rho k(\rho) \Theta(\rho) \int_{s_0}^s f(\sigma) d\sigma + p_K(\rho) 
$$ 
$$
= \rho k(\rho) \Theta(\rho) F_{s_0}(s) + p_K(\rho)
$$

and we can check that :

$$c^2(\rho,s) = \left[ k(\rho) + \rho k'(\rho) + k^2(\rho) \right] . \Theta(\rho) . F(s) + p'_K(\rho)$$
$$ = \left[ \frac{k'(\rho)}{k} + \frac{k(\rho) +1}{\rho} \right] (p - p_K(\rho)) + p'_K(\rho)$$

Also :
$$c_T^2(\rho,s) = c^2 - k^2 T C^v = $$
$$\left[k^2(\rho) + k(\rho) + \rho  k'(\rho)\right].\Theta(\rho) F(s) - k^2(\rho).\Theta(\rho).f(s).C^v(s) + p'_K(\rho) $$
$$ = \Theta(\rho) \left( \left[k^2(\rho)  + k(\rho) + \rho  k'(\rho)\right] F(s) - k^2(\rho).f(s).C^v(s) \right) + p'_K(\rho)$$
and the heat ratio is given by :
$$\gamma(\rho,s) = \frac{c^2}{c_T^2} =\frac{\left[k(\rho) (k(\rho) +1) + \rho  k'(\rho)\right] F(s)  + p'_K(\rho) /\Theta(\rho) } {\left[k(\rho) (k(\rho) +1) + \rho  k'(\rho)\right].F(s) - k^2(\rho).f(s).C^v(s) + p'_K(\rho) /\Theta(\rho) }$$
and 
$$
C_p (\rho,s)= \gamma(\rho,s)\, C_v(s).
$$

Finnaly, the complete form $e(\rho,s)$ is given by :
$$
e(\rho,s) = \Theta(\rho)F(s) + \int_{\rho_0}^{\rho} \frac{p_{K}(r)}{r^2}   dr.
$$
$$
e(\rho,s) = \Theta(\rho)F(s) +  e_{K}(\rho) 
$$

 
To summarize, the Mie-Grunenssein EOS is completely defined by the functions $k(\rho)$, $C^v(s)$ and $p_K(\rho)$ (or $e_K(\rho) = e_0+\int_{\rho_0}^{\rho} \frac{p_K(x)}{x^2} dx$) and a reference state ($\rho_0$, $T_0$, $s_0$, $p_0$, $e_0$).  
 
The dependance of $C^v$ on $s$ only implies that the Mie-Grunenssen EOS is not compatible with all models for $C^v$. A model of $C^v$ as fonction of $T$ is not compatibile with it. However the Debye model $C^v = C^v(\frac{T}{\Theta(\rho)})$ is compatible as $ \frac{T}{\Theta(\rho)}=f(s)$. The Einstein model, with a constant Enstein temperature $\Theta_E$, and $C^v$ of the form 
 $$C^v = 3R \left( \frac{\Theta_E}{T}\right)^2 \frac{e^{\frac{\Theta_E}{T}} }{\left( e^{\frac{\Theta_E}{T}} -1 \right)^2}$$
  is not. 
Two models that are compatible are for example 

$$C^v = 3R \left( \frac{\Theta}{T}\right)^2 \frac{e^{\frac{\Theta}{T}} }{\left( e^{\frac{\Theta}{T}} -1 \right)^2} ~~ \text(Enstein-Debye) $$
and 
$$C^v = 9R  \left( \frac{\Theta}{T}\right)^3 \int_0^{\frac{\Theta}{T}}  \frac{x^4 e^x}{(e^x -1)^2}dx  ~~  \text(Debye  model) $$

\subsection{Gruneissen and Stiffened gaz EOS}

We consider the special case of Mie-Gruneissen EOS where the Gruneisen parameter $k$ is supposed constant and where $C_v$ is supposed constant as well.

In this case 
$$\Theta(\rho) = T_0 \left( \frac{\rho}{\rho_0} \right) ^k$$
$$
c^2 = \frac{k+1}{\rho} (p -p_K(\rho)) + p'_K(\rho)
$$

$$f(s) = e^{\frac{s-s_0}{C^v}}$$.
$$F(s) = C^v \left(e^{\frac{s-s_0}{C^v}} - 1\right) = C^v (f(s) - 1) $$.
$$ c^2 = \Theta(\rho) k(k+1) C_v (f(s)-1)  + p'_K(\rho)$$
$$ c_T^2 = \Theta(\rho) \left(  k \, C_v \, f(s) - k(k+1) C_v \right) + p'_K(\rho)$$

and the heat ratio is given by :
$$\gamma(\rho,s) = \frac{c^2}{c_T^2} =\frac{ k(k+1)  C_v \, f(s) -  k(k+1)  C_v    + p'_K(\rho) /\Theta(\rho) } { k C_v f(s) - k(k+1) C_v + p'_K(\rho) /\Theta(\rho) }$$

It is worth noting that even in this simple case the heat ration $\gamma$ is not necessarly constant and equal to $k+1$ as usually assumed.
 
Only the special case $p'_K(\rho) =  k(k+1)C_v \Theta(\rho)$ so 
$$p_K(\rho) = p_0-  \rho_0 T_0 k C_v + \rho k C_v  \Theta(\rho) $$
gives 
$$
\gamma(\rho,s) = \gamma  = k+1 .
$$
and also implies the simple PVT relation :
$$
p = p_0 + \rho C_v T.
$$
and 
$$
e = e_0 + C_v T - \frac{p_0 - \rho_0 T_0 k C_v}{\rho}. 
$$

This case is exactly the Stiffened Gaz EOS \cite{Flatten} \cite{Menikoff}, with the Helmotz potential  :
$$
A(\rho,T) = C_v T \left(1 - ln (\frac{T}{T_0})+(\gamma -1) ln (\frac{\rho}{\rho_0}) \right) - s_0 \, T + \frac{p_\infty}{\rho} + e^* 
$$

We stress out that a relation of type $
p = \rho (\Gamma-1) (e-e_*) - \Gamma p_{\infty}
$, with constant $\Gamma$ is not suffiscent to have this simple PVT relation and to have a constant heat ratio $\gamma = \Gamma$, even if one supposes a constant $C_v$.

\section{Note on the implication of the second law of thermodynamics and stability}\label{2dPrinciple}

In the previous section we did only explicit the implications of the first law of thermodynamics and of the function state defintions, on the consistency of equations of state.

The second law of thermodynamics and its implication on mechanical and termal stability of the described matter,  can also be used to express more constraints on the `'degrees of freedom'' in an EOS relation. 

First we consider a $e(v=\frac{1}{\rho},s)$ representation and compute
$$
x = \frac{\partial^2 e}{\partial v^2} =\frac{c^2}{v^2} = \rho^2 c^2  
$$

$$
y = \frac{\partial^2 e}{\partial s^2} =  \frac{T}{C_v} 
$$

$$
z = \frac{\partial^2 e}{\partial s \partial v} =  - \frac{kT}{v} 
$$

$$
x\,y - z^2 =   \frac{T}{C_v}   \frac{c^2 - k^2T C_v}{v^2}  = \frac{T}{C_v} \frac{c_T^2 }{v^2}
$$

Mechanical statibility implies for instance (See \cite{Prigogine} page 209) :
 $$
 c_T^2>0,
 $$
while thermal stability implies
$$
C_v > 0.
$$

A more general condition for statbilty of equilibrium states, implied by the second law of thermodynamics is the convexity of $e(v,s)$ (see the introduction of \cite{Israel}). Hence if the hessien matrix of $e(\rho,v)$ is definite positive, wich boild down to the same identities $C_v>0$, $c^2>0$ and $c_T^2 >0$.

This also implied $\gamma >0$ and  $Cp >0$ , and by Mayer relation   $Cp-Cv =  T \alpha^2 c_T^2 > 0$  we have $\gamma >1$.

\section{Conclusion}
We presented a way to construct a complete set of thermodynamic relations and reviewed the thermodynamic consistency of many classical Equaion of States. The approach presented is general and can be used when `inventing' new equations of state in research work or when fitting an EOS and thermodynamic coeffiscients with laboratory data.  The approch allows to explicit `'the degrees of freedom`' for each EOS to avoid violating the thermodynamic consistency. As example, we explained by an elementary approach why many classical equations of state, such as stiffened gaz (with constant heat ratio $\gamma$) can not be generalized by making $C_v$ dependent on the temperature $T$ without violating thermodynamical consistency.



\end{document}